\documentclass[12pt]{article}
\usepackage{amssymb}
\newcommand{\C}{\mathbb C }
\newcommand{\N}{\mathbb N}
\newcommand{\R}{\mathbb R}
\def\kasten{$~~\mbox{\hfil\vrule height6pt width5pt depth-1pt}$ }

\newtheorem{theorem}{Theorem}[section]
\newtheorem{definition}[theorem]{Definition}
\newtheorem{lemma}[theorem]{Lemma}
\newtheorem{remark}[theorem]{Remark}
\newtheorem{condition}[theorem]{Condition}
\newtheorem{corollary}[theorem]{Corollary}
\begin{document}

\title{A Characterisation of Locality in Momentum Space}
\author{Hanno Gottschalk\\
Universit\`a di Roma ``La Sapienza'',\\ Dipartimento di
Mathematica,\\ Piazz. Aldo Moro 2, I-00185 Roma, Italy.\\gottscha@giove.mat.uniroma1.it}
\maketitle
\begin{abstract}
It is proved that a Poincar\'e invariant Wightman function which
fulfils the spectral property and can be defined at sharp times is
local if and only if the integration over both the energy
variables of a commutator in momentum space is a polynomial in the
momentum conjugated to the spacial difference variable of the
commutator with distributional coefficients depending on the
remaining energy and momentum variables. Using this
characterisation of locality in momentum space, the locality of a
sequence of Wightman functions with nontrivial scattering
behaviour (associated to some quantum field in indefinite metric)
can be proved by explicit calculations. We compare the above
characterisation of locality with the classical integral
representation method of Jost, Lehmann and Dyson.
\end{abstract}

{{\bf AMS
subject classification (1991):}{81T05}

\section*{Introduction}

The vacuum expectation values (Wightman functions) of a local
relativistic quantum field contain all the general and specific
information of the underlying quantum field theory (QFT).

The formulation of the general (axiomatic) properties of
(truncated) Wightman functions given in the classical literature
\cite{Jo,SW} requires expressions for the Wightman functions in
position space and in momentum space. In applications however, the
passage from one picture to the ``dual picture'' by the Fourier
transform requires quite nontrivial calculations. It would
therefore be convenient to get formulations for all axiomatic
properties in only one of these ``pictures''.

It seems to be somehow more natural to study (truncated) Wightman
functions in momentum space than in position space, since in
momentum space there are direct formulations for the spectral
property and cluster property in terms of the support of the
(truncated) Wightman functions. Furthermore, Poincar\'e
invariance, mass shell singularities leading to nontrivial
scattering behaviour etc. can be directy identified from
(truncated) Wightman functions in momentum space. It therefore is
an interesting problem to get a description of locality in
momentum space.

A characterisation of locality in momentum space has been given by
Jost, Lehmann and Dyson (JLD) in the form of an integral
representation of causal commutators \cite{Dy,JL}. While this
integral representation is a very powerful tool to investigate
structural properties (e.g. analyticity) of causal commutators in
momentum space, it seems non-trivial to decide the question of the
locality of a given Wightman function in momentum space on the
basis of the JLD--representation, since this amounts more or less
to the calculation of the inverse of an integral transform. Also,
as we shall demonstrate in this article, the JLD--representation
tacitly requires some regularity assumptions, which rule out some
cases of commutators of local Wightman functions.

Here we give a new characterisation of locality in momentum space
which seems to give a very straight forward criterium to check
locality in momentum space. This is being illustrated by a
concrete application to a physically nontrivial situation. We also
apply our method to the JLD--representation and we provide
examples which show that our characterisation method for causal
commutators goes properly beyhond the result of \cite{Dy,JL}.

The article is organised as follows: After collecting some
notations and definitions in Section 1 we propose such a criterion
for locality in momentum space (Section 2). We then in Section 3
apply this criterion  to a sequence of truncated Wightman
functions (associated to a physically nontrivial QFT in
``indefinite metric'', cf. \cite{AG,AGW2,AGW3,Go}) which were
constructed in \cite{AGW1} (see also the references in this
article). The proof of locality obtained along this line is much
shorter and simpler than the proof of the same statement based on
Euclidean QFT and analytic continuation which is given in
\cite{AGW1}. Finally, in Section 4 we compare our method with the
integral representation method of JLD \cite{Dy,JL}.

\section{Double integrals over the energy variables of a commutator}

Let $\R^d$, $d\geq 2$, be the $d$-dimensional Minkowski space time
with inner product $x\cdot y=x^0y^0-\vec x\cdot\vec y, x=(x^0,\vec
x),y=(y^0,\vec y)\in \R\times\R^{d-1}=\R^d$. For $x\cdot x$ we
also use the expression $x^2$. We denote the forward (backward)
light cone $\{x\in\R^d:x^2>0,x^0>0 (x^0<0)\}$ by $ V_0^+(V_0^-)$
and $\bar V_0^+ (\bar V_0^-)$ is the closed forward
(backward)light cone.

We deal with the $n$-point vacuum expectation values of a QFT with
$N$ species of quantum fields labeled by indices
$\kappa_l,l=1,\ldots,n,$ which transform covariantly under spin
representations $T_{\kappa_l}$ of the covering group of the
Lorentz group $\tilde L^\uparrow_+$ over $\R^d$ with finite
dimensional spin space $E_{\kappa_l}$. The index
$\nu_l=1,\ldots,\mbox{dim}_{\C}E_{\kappa_l}$ is the spin-index of
the quantum field of species $\kappa_l$.

By ${\cal S}_n$ we denote the space of Schwartz testfunctions over
$\R^{dn}$ with values in $E^{\otimes n}$ where $E=\oplus
_{\kappa=1}^ NE_\kappa$. A (truncated) $n$-point Wightman function
$W_n$ is an element in the topological dual space ${\cal S}_n'$ of
${\cal S}_n$.

The components of the Wightman functions are denoted by \linebreak
$W_n^{(\kappa_1,\ldots,\kappa_n)\nu_1\cdots\nu_n}
(x_1,\ldots,x_n)$ and their Fourier transform is defined as
\begin{eqnarray}
\label{0eqa}
 \hat W_n^{(\kappa_1,\ldots,\kappa_n)\nu_1\cdots\nu_n}(k_1,\ldots,k_n)
&=&(2\pi)^{-nd/2}\int_{\R^{dn}}e^{-i(k_1\cdot x_1+\cdots+k_n\cdot x_n)}\nonumber \\
&\times& \!\! W_n^{(\kappa_1,\ldots,\kappa_n)\nu_1\cdots\nu_n}(x_1,\ldots,x_n)~dx_1\cdots dx_n.\nonumber\\
\end{eqnarray}
Here the integral on the right hand side has to be understood in
the sense of the Fourier transform of tempered distribuions, cf.
\cite{Co}. In the following, we also use the symbol ${\cal F}$ for
the Fourier transform and we denote the inverse Fourier transform
by $\bar {\cal F}$.

 We assume that $W_n$ fulfils the \underline{spectral property} $\mbox{supp}~\hat W_n\subseteq \{ (k_1,\linebreak \ldots,k_n)\in \R^{dn}:\sum_{l=j}^nk_l\in \bar V_0^+, j=1,\ldots,n-1\}$ and the property of \underline{Poincar\'e invariance}
\begin{eqnarray}
\label{1eqa}
&&\prod_{l=1}^nT^{\nu'_l}_{\kappa_l,\nu_l}(\Lambda) W_n^{(\kappa_1,\ldots,\kappa_n)\nu_1\cdots\nu_n}(\Lambda^{-1}(x_1-a),\ldots,\Lambda^{-1}(x_n-a))\nonumber\\
&=&
W_n^{(\kappa_1,\ldots,\kappa_n)\nu_1'\cdots\nu_n'}(x_1,\ldots,x_n)~~\forall
\Lambda \in \tilde L_+^\uparrow, ~a\in\R^d,
\end{eqnarray}
where we applied the Einstein convention of summation (ECS), i.e.
any spin index $\nu_l$ is summed up over
$1,\ldots,\mbox{dim}_{\C}E_{\kappa_l}$.

The desired (anti-) commutation relations of a field of type
$\kappa$ with a field of type $\kappa'$ are being fixed by a
symmetric $N\times N$-matrix $\sigma$,
$\sigma^{\kappa,\kappa'}=\pm 1$, $\kappa,\kappa'=1,\ldots,N$.

By definition, a (truncated) Wightman function $W_n$ is
\underline{local} (w.r.t. $\sigma$) if and only if for
$x_j-x_{j+1}$ is space-like ( i.e. $(x_j-x_{j+1})^2<0$) we get
\begin{eqnarray}
\label{2eqa} &&W_{n}^{(\kappa_1,\ldots,\kappa_n)
\nu_1\cdots\nu_n}(x_1,\ldots,x_n)=\nonumber \\
&&\sigma^{\kappa_j,\kappa_{j+1}}
W_{n}^{(\kappa_1,\ldots,\kappa_{j+1},\kappa_j,\ldots,\kappa_n)
\nu_1\cdots\nu_{j+1}\nu_j\cdots\nu_n}(x_1,\ldots,
x_{j+1},x_j,\ldots,x_n).
\end{eqnarray}

For $j=1,\ldots,n-1$, we define the distribution
$W_{n,[,]_j}\in{\cal S}_n'$ by
\begin{eqnarray}
\label{3eqa}
&&W_{n,[,]_j}^{(\kappa_1,\ldots,\kappa_n)\nu_1\cdots\nu_n}(x_1,\ldots,x_n)\nonumber\\
&=&
W_n^{(\kappa_1,\ldots,\kappa_n)\nu_1\cdots\nu_n}(x_1,\ldots,[x_j,x_{j+1}],\ldots,x_n)\nonumber\\
&=&W_n^{(\kappa_1,\ldots,\kappa_j,\kappa_{j+1},\ldots\kappa_n)\nu_1\cdots\nu_j\nu_{j+1}\cdots\nu_n}(x_1,\ldots,x_j,x_{j+1},\ldots,x_n)\nonumber\\
&-&\sigma^{\kappa_j,\kappa_{j+1}}W_n^{(\kappa_1,\ldots,\kappa_{j+1},\kappa_{j},\ldots\kappa_n)\nu_1\cdots\nu_{j+1}\nu_{j}\cdots\nu_n}(x_1,\ldots,x_{j+1},x_{j},\ldots,x_n)\nonumber\\
\end{eqnarray}

Let $\varphi$ be a symmetric, real Schwartz function on $\R$ with
support in $[-1,1]$ and $\int_{\R}\varphi\, dx =1$. We set
$\varphi_\epsilon(x)=\varphi(x/\epsilon)/\epsilon$. We define
$W_{n,j,\epsilon}$ as the distribution $W_{n}$ convoluted with
$\varphi_\epsilon$ in each of the arguments $x_j^0,x_{j+1}^0$. The
distribution
$W_{n,j,\epsilon}^{(\kappa_1,\ldots,\kappa_n)\nu_1\cdots\nu_n}(x_1,\ldots,x_n)$
 thus is a smooth and polynomially bounded function in the arguments $x_j^0,x_{j+1}^0$, provided it is smeared out
with some testfunction in the remaining arguments
$x_1,\ldots,x_{j-1},\vec x_j,\vec x_{j+1},x_{j+2},\ldots,x_n$.
Furthermore, $\lim_{\epsilon\to+0}W_{n,j,\epsilon}=W_n$ for
$j=1,\ldots,n-1$. For $W_{n,[,]_j,j,\epsilon}$ we write
$W_{n,[,]_j,\epsilon}$.

Frequently we need the testfunction spaces ${\cal
S}_{n,j},j=1,\ldots,n-1$ which are the spaces of Schwartz
functions of the arguments $x_1,\ldots,\linebreak x_{j-1},\vec
x_j,\vec x_{j+1},x_{j+2},\ldots,x_n$ with values in $E^{\otimes
n}$ with the Schwartz topology and their topological dual spaces
${\cal S}_{n,j}'$. By $W_{n,j,\epsilon}(s,t)\in {\cal S}_{n,j}'$
we denote the distribution which is defined by
\begin{eqnarray}
\label{4eqa} W_{n,j,\epsilon}(s,t)(f)
&=&\int_{\R^{dn}} W_n^{(\kappa_1,\ldots,\kappa_n)\nu_1\cdots\nu_n}(x_1,\ldots,x_n)\nonumber\\
&\times&f_{(\kappa_1,\ldots,\kappa_n)\nu_1\cdots\nu_n}(x_1,\ldots,x_{j-1},\vec x_j,\vec x_{j+1},x_{j+2},\ldots,x_n)\nonumber\\
&\times&
\varphi_\epsilon(s-x_j^0)\varphi_\epsilon(t-x_{j+1}^0)~dx_1\cdots
dx_n
\end{eqnarray}
where we applied the ECS to the indices
$\kappa_l,\nu_l,l=1,\ldots,n$.

In order to formulate our condition of locality, we need the
following rather weak and technical restriction on the Wightman
functions $W_n$:

\begin{condition}
\label{1cond} {\rm We say that $W_n$ fulfils the weak time zero
field condition, if
$W_{n,j}(0,0)=\lim_{\epsilon\to+0}W_{n,j,\epsilon}(0,0)$ exists in
${\cal S}_{n,j}'$ for $j=1,\ldots,n-1$. }
\end{condition}

\begin{remark}
\label{1rem} {\rm (i) Condition \ref{1cond} follows from the
existence of sharp time fields $\phi(\delta_{t}\otimes f)$ where
$f\in{\cal S}_1(\R^{d-1},E)$ and $\delta_{t}(x^0)=\delta(x^0-t)$
at time zero. However, no precise assumptions are made on the
domains of definition of such quantum fields, and we therefore
labeled it with the adjective "weak".

\noindent (ii) Of course, if the weak time zero field condition
holds for $W_n$, then it also holds for $W_{n,[,]_j}$ and we get a
distribution $W_{n,[,]_j}(0,0)$ in ${\cal S}_{n,j}'$.

\noindent (iii) Formally we get the following expression for the
Fourier transform (in ${\cal S}_{n,j}'$) $\hat W_{n,[,]_j}(0,0)$
of $W_{n,[,]_j}(0,0)$:
\begin{eqnarray}
\label{5eqa}
&&\hat W_{n,[,]_j}(0,0)(k_1,\ldots,k_{j-1},\vec k_j,\vec k_{j+1},k_{j+1},\ldots,k_n)\nonumber\\
&=&\int_{\R}\int_{\R}\hat
W_n(k_1,\ldots,[k_j,k_{j+1}],\ldots,k_n)~dk_j^0dk_{j+1}^0
\end{eqnarray}
}
\end{remark}

For $j=1,\ldots,n-1$ we define $\xi_+=(x_j+x_{j+1})/2$ and
$\xi_-=(x_j-x_{j+1})/2$. The variables conjugated to $x_1,\ldots,
x_{j-1},\xi_-,\xi_+,x_{j+1},\ldots,x_n$ under the Fourier
transform are  $k_1,\ldots, k_{j-1},q_-,q_+,k_{j+1},\ldots,k_n$
with $q_\pm =(k_j\pm k_{j+1})/2$. We define another testfunction
space ${\cal S}_{n,j,+}$ as the space of Schwartz functions on
$\R^{d(j-1)}\times\R^{d-1}\times\R^{d(n-j-2)}$ with values in
$E^{\otimes n}$. Given $f\in{\cal S}_{n,j,+}$ we define the
tempered distribution in the argument $\vec q_-$, $\hat
W_{n,[,]_j}(0,0)(f)(\vec q_-),$ as
\begin{eqnarray}
\label{6eqa} \hat W_{n,[,]_j}(0,0)(f)(\vec q_-)=
\int_{\R^{d(n-2)+(d-1)}} \hat W_{n,[,]_j}^{(\kappa_1,\ldots,\kappa_n)\nu_1\cdots\nu_n}(0,0)(k_1,\ldots,&&\nonumber\\
k_{j-1},\vec q_++\vec q_-,\vec q_+-\vec q_-,k_{j+2},\ldots,k_n)&&\nonumber\\
\times f_{(\kappa_1,\ldots,\kappa_n)\nu_1\cdots\nu_n}(k_1,\ldots,k_{j-1},\vec q_+,k_{j+2},\ldots,k_n))&&\nonumber\\
\times~dk_1\cdots dk_{j-1}d\vec q_+dk_{j+2}\cdots dk_n,&&
\end{eqnarray}
where we again used the ECS. By an analogous formula we
define\linebreak $W_{n,[,]_j}(0,0)(f)(\vec \xi_-)$ for $f\in{\cal
S}_{n,j,+}$. If $f\in{\cal S}_{n,j}$ we then define
$f_{+}(\vec\xi_-)\in{\cal S}_{n,j,+}$ for $\vec\xi_-\in\R^{d-1},
j=1,\ldots,n-1$ as
\begin{eqnarray}
\label{7eqa}
&&f_+(\vec \xi_-)(x_1,\ldots,x_{j-1},\vec\xi_+,x_{j+2},\ldots,x_n) \nonumber \\
&=&f(x_1,\ldots,x_{j-1},\vec \xi_++\vec \xi_-,\vec
\xi_+-\vec\xi_-,x_{j+2},\ldots,x_n) .
\end{eqnarray}
Then, we get
\begin{equation}
\label{8eqa} W_{n,[,]_j}(0,0)(f)=\int_{\R^{d-1}}
W_{n,[,]_j}(0,0)(f_{+}(\vec \xi_-))(\vec
\xi_-)~d\vec\xi_-~~\forall f\in{\cal S}_{n,j}.
\end{equation}
where the  integral is a symbolic ("distributional") integral.

\section{The main theorem}

We have now finished the preparations for the formulation of the
following criterion for locality in momentum space:

\begin{theorem}
\label{1theo} Let $W_n\in{\cal S}_n'$ be a Poincar\'e invariant
distribution which fulfils the spectral property and the weak time
zero field condition \ref{1cond}. Then $W_n$ is local if  and only
if $\hat W_{n,[,]_j}(0,0)(f)(\vec q_-)$ is a polynomial in $\vec
q_-$ for all $f\in{\cal S}_{n,j,+}$.
\end{theorem}

\noindent {\bf Proof.}
$\Rightarrow$ : Let $j\in\{1,\ldots,n-1\}$ and $f\in{\cal
S}_{n,j,+}$ be fixed.

We note that the polynomials are the Fourier transforms of the
complex linear combinations of the delta distribution in the point
$0\in\R^{(d-1)}$ and it's partial derivatives.

Thus, the distribution $\hat W_{n,[,]_j}(0,0)(f)(\vec q_-)$ is a
polynomial in $\vec q_-$ if and only if its inverse Fourier
transform $\bar{\cal F}_{\vec q_-}(\hat W_{n,[,]_j}(0,0)(f))(\vec
\xi_-)$ is a linear combination of the delta distribution in the
point $0\in\R^{(d-1)}$ and its partial derivatives.

By \cite{Co} p. 56 we know that the distributions with support in
$\{0\}\subseteq\R^d$ are just given by the linear combination of
the delta distribution in the point $0\in\R^{(d-1)}$ and its
partial derivatives. It is thus equivalent to show that
$\mbox{supp}\,\bar{\cal F}_{\vec q_-}(\hat
W_{n,[,]_j}(0,0)(f))(\vec \xi_-)\subseteq \{0\}$. Let
$B_{\epsilon_1}(0)\subseteq \R^{d-1}$ be the ball of radius
$\epsilon_1$ with center $0$. We have to show that for any
$\epsilon_1>0$ and Schwartz function $h(\vec \xi_-)$ with support
in $\R^{d-1}-B_{\epsilon_1}(0)$ we have
$$
\int_{\R^{d-1}}\bar{\cal F}_{\vec q_-}(\hat
W_{n,[,]_j}(0,0)(f))(\vec \xi_-)h(\vec \xi_-)~d\vec \xi_-=0.
$$
Rewriting this equation in terms of $W_{n,[,]_j}$ we get for the
left hand side
\begin{eqnarray*}
&&\lim_{\epsilon \to +0}\int_{\R^{dn}}W_{n,[,]_j}^{(\kappa_1,\ldots,\kappa_n)\nu_1\cdots\nu_n}(x_1,\ldots,x_n)\\
&&(\bar {\cal F}f)_{(\kappa_1,\ldots,\kappa_n)\nu_1\cdots\nu_n}(x_1,\ldots,x_{j-1},\vec x_j+\vec x_{j+1},x_{j+2},\ldots,x_n)\\
&&\times h((\vec x_j-\vec
x_{j+1})/2)\varphi_\epsilon(x_j^0)\varphi_{\epsilon}(x_{j+1}^0)~dx_1\cdots
dx_n.~~~~(ECS)
\end{eqnarray*}
We note that for $0<4\epsilon<\epsilon_1$ the support of the
function $h((\vec x_j-\vec x_{j+1})/2)
\times\varphi_\epsilon(x_j^0) \varphi_{\epsilon}(x_{j+1}^0)$ is
contained in $\{ (x_j,x_{j+1})\in\R^d\times\R^d:
(x_j-x_{j+1})^2<0\}$. Thus, by the locality of $W_n$, the
"integral" in the above expression is zero for such $\epsilon$ and
thus the limit is zero.

$\Leftarrow$: First we fix some notations and recall some results
of axiomatic QFT following \cite{SW}.

By the spectral property, the Wightman functions
$W_n^{(\kappa_1,\ldots,\kappa_n)\nu_1,\ldots,\nu_n}\linebreak(x_1,\ldots,x_n)$
are boundary values of holomorphic functions\linebreak ${\cal
W}_n^{(\kappa_1,\ldots,\kappa_n)\nu_1,\ldots,\nu_n}(z_1,\ldots,z_n)$,
$z_j=x_j+iy_j,j=1,\ldots,n$ which are analytic in the tube ${\cal
T}_n=\R^n+i\Gamma_n$ with $\Gamma_n=\{
(y_1,\ldots,y_n)\in\R^{dn}:y_j-y_{j+1}\in V_0^-\}$. Let $\tilde
L_+(\C^d)$ be the (covering group of the) proper complex Lorentz
group. Then, by Poincar\'e invariance,${\cal
W}_n^{(\kappa_1,\ldots,\kappa_n)\nu_1,\ldots,\nu_n}(z_1,\ldots,\linebreak
z_n)$ has a single valued extension to the extended tube ${\cal
T}_n'=\tilde L_+(\C^d)\cdot{\cal T}_n$ (here the dot stands for
the diagonal action of $\tilde L_+(\C^d)$ on $\C^{dn}$). The real
points in ${\cal T}'_n$ are the so called Jost points, i.e. the
points $\{
(x_1,\ldots,x_n)\in\R^{dn}:(\sum_{j=1}^{n-1}\lambda_j(x_j-x_{j+1}))^2<0\forall
\lambda_j\geq0,\sum_{j=1}^{n-1}\lambda_j>0\}$.

Similarly, for $\pi\in\mbox{Perm}(n)$ the permuted Wightman
function\linebreak $W_{n,\pi}^{(\kappa_1,\ldots,\kappa_n)
\nu_1\cdots,\nu_n}(x_1,\ldots,x_n)=
W_n^{(\kappa_{\pi_1},\ldots,\kappa_{\pi_n})\nu_{\pi_1},\ldots,\nu_{\pi_n}}(x_{\pi_1},\ldots,x_{\pi_n})$
is \linebreak the  boundary value of a holomorphic function
 ${\cal W}_{n,\pi}^{(\kappa_1,\ldots,\kappa_n)
\nu_1\cdots,\nu_n}(z_1,\ldots,\linebreak z_n)$ defined on the
extended tube ${\cal T}_{n,\pi}'=p(\pi)\cdot{\cal T}_n'$ where $p$
denots the action of the permutation group on $\C^{dn}$.

For $j=1,\ldots,n-1$ let $(j,j+1)\in\mbox{Perm}(n)$ denote the
transposition of $j$ and $j+1$. Then, ${\cal T}_n'$ and ${\cal
T}_{n,(j,j+1)}'$ have a nonempty real
 open intersection ${\cal N}_{n,(j,j+1)}$. Furthermore, since the action of $\tilde L^\uparrow_+$ maps real points to real points and leaves ${\cal T}_n'$ and ${\cal T}_{n,(j,j+1)}'$
invariant, ${\cal N}_{n,(j,j+1)}$ is invariant under the action of
$\tilde L^\uparrow_+$. Similarly, ${\cal N}_{n,(j,j+1)}$ is
invariant under (real) translations. If
\begin{equation}
\label{9eqa} {\cal W}_{n}^{(\kappa_1,\ldots,\kappa_n)
\nu_1\cdots\nu_n}(z_1,\ldots,z_n)=\sigma^{\kappa_j,\kappa_{j+1}}{\cal
W}_{n,(j,j+1)}^{(\kappa_1,\ldots,\kappa_n)
\nu_1\cdots,\nu_n}(z_1,\ldots,z_n)
\end{equation}
holds on ${\cal N}_{n,(j,j+1)}$, then ${\cal
W}_{n}^{(\kappa_1,\ldots,\kappa_n)
\nu_1\cdots,\nu_n}(z_1,\ldots,z_n)$ and \linebreak ${\cal
W}_{n,(j,j+1)}^{(\kappa_1,\ldots,\kappa_n)
\nu_1\cdots,\nu_n}(z_1,\ldots,z_n)$ have single valued
continuation on ${\cal T}'_n\cup{\cal T}_{n,(j,j+1)}'$ and the
relation (\ref{9eqa}) holds on this domain.

Since the transpositions generate the group of permutations, it is
sufficient to prove (\ref{9eqa}) for $j=1,\ldots,n-1$ in order to
obtain analytic functions ${\cal
W}_{n,\pi}^{(\kappa_1,\ldots,\kappa_n)\nu_1,\cdots,\nu_n}(z_1,\ldots,z_n)$
defined on the permuted extended tube ${\cal
T}_n^{p.e.}=\cup_{\pi\in{\rm Perm}(n)}{\cal T}_{\pi,n}'$, s.t. the
relation (\ref{9eqa}) and related relations between these
functions hold on ${\cal T}_n^{p.e.}$. If this is true, then, by a
general theorem of R. Jost \cite{Jo} p. 83, $W_n$ is local.

It is thus sufficient to prove Equation (\ref{9eqa}) on ${\cal
N}_{n,(j,j+1)}$ for $j=1,\ldots,n-1$. Since the points in ${\cal
N}_{n,(j,j+1)}$ are real, this equation can be written in terms of
the Wightman functions themselves, i.e. we have to show that
\begin{eqnarray}
\label{10eqa} &&W_{n}^{(\kappa_1,\ldots,\kappa_n)
\nu_1\cdots\nu_n}(x_1,\ldots,x_n)\nonumber \\
&=&\sigma^{\kappa_j,\kappa_{j+1}}
W_{n}^{(\kappa_1,\ldots,\kappa_{j+1},\kappa_j,\ldots,\kappa_n)
\nu_1\cdots\nu_{j+1}\nu_j\cdots\nu_n}(x_1,\ldots,
x_{j+1},x_j,\ldots,x_n)\nonumber \\
&&
\end{eqnarray}
holds for $(x_1,\ldots,x_n)\in{\cal N}_{n(j,j+1)}$. We note that
the above equation is a relation between real analytic functions
and therefore no smearing in the variables $x_1,\ldots,x_n$ is
required in order to make it rigorous. We can thus fix
$(x_1,\ldots,x_n)$.

Since $(x_1,\ldots,x_n)\in {\cal N}_{n,(j,j+1)}$,
$(x_1,\ldots,x_n)$ is a Jost point and we get that $x_j-x_{j+1}$
is space like. Thus there exists a Lorentz transformation
$\Lambda\in L^\uparrow_+$ s.t. $\Lambda^{-1}$ maps $x_j-x_{j+1}$
to the hyperplane $\{0\}\times\R^{d-1}$. Equation (\ref{10eqa})
thus is equivalent with
\begin{eqnarray*}
&&\prod_{l=1}^n
T_{\kappa_l,\nu_l}^{\nu_l'}(\Lambda)W_{n}^{(\kappa_1,\ldots,\kappa_n)
\nu_1\cdots\nu_n}(\Lambda^{-1}x_1,\ldots,\Lambda^{-1}x_n) \\
&=&\sigma^{\kappa_j,\kappa_{j+1}} \prod_{l=1}^n
T_{\kappa_l,\nu_l}^{\nu_l'}(\Lambda)
W_{n}^{(\kappa_1,\ldots,\kappa_{j+1},\kappa_j,\ldots,\kappa_n)
\nu_1\cdots\nu_{j+1}\nu_j\cdots\nu_n}(\Lambda^{-1}x_1,\\
&&\ldots,\Lambda^{-1}
x_{j+1},\Lambda^{-1}x_j,\ldots,\Lambda^{-1}x_n)
\end{eqnarray*}
where we applied the ECS to the indices $\nu_l$. It is therefore
sufficient to prove Equation (\ref{10eqa}) for the points
$(x_1,\ldots x_n)\in{\cal N}_{n,(j,j+1)}$ replaced by the points
$(x_1',\ldots,x_n')= (\Lambda^{-1}x_1,\ldots,\Lambda^{-1}x_n)\in
{\cal N}_{n,(j,j+1)}$, where ${x'}_j^{0}={x'}_{j+1}^{0}$.
Furthermore, by the translation invariance of the Wightman
functions and the translation invariance on ${\cal
N}_{n,(j,j+1)}$, this is equivalent with Equation (\ref{10eqa})
for the points $(x_1,\ldots,x_n)$ replaced with the points
$(x_1'',\ldots,x_n'')\in{\cal N}_{n,(j,j+1)}$ where
$x''_l=({x'}_l^{0}-{x'}_j^{0},\vec x_l'), l=1,\ldots,n$. It is
thus sufficient to prove Equation (\ref{10eqa}) for points
$(x_1,\ldots,x_n)\in{\cal N}_{n,(j,j+1)}$ with
$x_j^0=x_{j+1}^0=0$. We denote the set of these points by ${\cal
N}_{n,(j,j+1)}^0$. Since ${\cal N}_{n,(j,j+1)}$ is open in
$\R^{dn}$, we can also consider ${\cal N}_{n,(j,j+1)}^0$ as an
open subset of
$\R^{d(j-1)}\times\R^{d-1}\times\R^{d-1}\times\R^{d(n-j-2)}$. That
Equation (\ref{10eqa}) holds on ${\cal N}_{n,(j,j+1)}^0$ thus is
equivalent to
$$
W_{n,[,]_j}(0,0)(f)=0~~\forall f\in {\cal
S}_{n,j},~~\mbox{supp}~f\subseteq {\cal N}_{n,(j,j+1)}^0.
$$
By equation (\ref{8eqa}) it is sufficient to show that the
distribution (in $\vec \xi_-$)\linebreak
$W_{n,[,]_j}(0,0)(f_+(\vec\xi_-))(\vec\xi_-)=0$ for $f\in{\cal
S}_{n,j}$ with $\mbox{supp }f\subseteq {\cal N}_{n(j,j+1)}^0$. As
in the first part of the proof we get from the fact that $\hat
W_{n,[,]_j}(0,0)\linebreak ({\cal F}f_+(\vec\xi_-))(\vec q_-)$ is
a polynomial in $\vec q_-$ (here the Fourier transform ${\cal F}$
is the Fourier transform in ${\cal S}_{n,j,+}$ and $\vec\xi_-$ is
a fixed parameter)
 that $W_{n,[,]_j}(0,0)(f(\vec \xi_-))(\vec \xi_-)$ has support concentrated in $\{0\}\subseteq \R^{d-1}$. We note that for $(x_1,\ldots,x_{j-1},\vec x_j,\linebreak\vec x_{j+1},x_{j+2},\ldots,x_n)
\in {\cal N}_{n,(j,j+1)}^0$ we have $\vec x_j\not =\vec x_{j+1}$,
since ${\cal N}_{n,(j,j+1)}^0$ consists of Jost points which
implies $(0,\vec x_j-\vec x_{j+1})^2<0$. Thus, $f_+(\vec\xi_-)=0$
on a neighbourhood of $0\in\R^{d-1}$. Consequently,
$W_{n,[,]_j}(0,0)(f_+(\vec\xi_-))(\vec\xi_-)=0$ holds on this
neighbourhood and therefore holds everywhere.\kasten

\section{Application: Locality of the structure functions}

As an immediate consequence of Theorem \ref{1theo}, one obtains
the locality of the two point function $\hat
W_2(k_1,k_2)=\delta_m^-(k_1)\delta(k_1+k_2)$ of the free field of
mass $m$ as follows (here $\delta^\pm_m(k)=\theta(\pm
k^0)\delta(k^2-m^2)$ with $\theta$ beeing the Heaviside function):
It is well-known, that spectrality and Poincar\'e invariance hold
for this distribution and that also the weak time zero field
condition holds. Thus, the following short calculation suffices to
prove locality:
\begin{eqnarray}
\int_{\R}\int_{\R}\hat W_{2,[,]_1}(k_1,k_2)dk_1^0dk_2^0 &=&\int_{\R}\int_{\R}\left(\delta^-_m(k_1)-\delta^-_m(k_2)\right)\nonumber \\
\times\delta(k_1+k_2)dk_1^0dk_2^0 &=& \left({1\over 2
\omega_1}-{1\over 2 \omega_2}\right)\delta(\vec k_1+\vec k_2)=0
\end{eqnarray}
where $\omega_j=\sqrt{|\vec k_j|^2+m^2},~j=1,2$.

But in this section we want to show that Theorem \ref{1theo} is
useful especially in physically nontrivial situations. To do this,
we define a sequence of truncated Wightman functions, called the
\underline{structure func-} \underline{tions}, which play a
crucial r\^ole in the construction of quantum fields in indefinite
metric with nontrivial scattering behaviour given in
\cite{AG,AGW1,AGW2,AGW3,Go}. In fact,  the scattering amplitudes
associated to the structure functions just consist of ``on shell''
and energy-momentum conservation terms. Using the characterisation
of locality in momentum space, we derive the locality of these
truncated Wightman functions. This result is implicitly already
contained in \cite{AGW1}. However, the proof given there uses
``Euclidean'' methods and analytic continuation and is much longer
than the proof we present here.

For $n\in\N, n\geq 3,$ let $m_1,\ldots,m_N\in\R^n, m_\kappa>0$ if
$d=2,3$ and $m_\kappa\geq 0$ for $d\geq 4,~\kappa=1,\ldots,N$. Let
$\bar{\kappa}=(\kappa_1,\ldots,\kappa_n)$. We then define the
distributions
\begin{equation}
\label{12eqa} \hat G_{n,\bar{ \kappa}}(k_1,\ldots,k_n)=
\left\{\sum_{j=1}^n\prod_{l=1}^{j-1}\delta^-_{m_{\kappa_l}}(k_l)
{1\over
k^2_j-m_{\kappa_j}^2}\prod_{l=j+1}^n\delta^+_{m_{\kappa_l}}(k_l)\right\}\delta
(\sum_{l=1}^nk_l)
\end{equation}
Here the singularities $1/(k_j^2-m_{\kappa_j}^2)$ have to be
understood in the sense of Cauchy's priciple value, cf. \cite{Co}
p.44.

\begin{definition}
\label{2def} {\rm For $n\geq3$ we define the structure function
$G_n$ as the inverse Fourier transform of the distribution $\hat
G_n$ given by $\hat G_n =\sum_{\kappa_1,\ldots,\kappa_n=1}^N\hat
G_{n,\bar{\kappa}}$. \rm}
\end{definition}

The distributions $\hat G_n$ are manifestly Poincar\'e invariant.
Furthermore, they fulfil the spectral condition, which can be
proved as follows: Let $(k_1,\ldots,k_n)$ be in the support of the
$j$-th summand of $\hat G_{n,\bar \kappa}$. Then
$\sum_{l=r}^nk_l\in \bar V_0^+$ for $ r=j+1,\ldots,n-1$ since
$k_l\in \bar V_0^+$ for $l=r,\ldots,n-1$. If $1\leq r\leq j$ then
$\sum_{l=r}^nk_l=-\sum_{l=1}^{r-1}k_l\in\bar V_0^+$ since $k_l\in
-\bar V_0^+$ for $l=1,\ldots,j-1$. In order to apply Theorem
\ref{1theo}, it remains to prove the following lemma:

\begin{lemma}
\label{1lem} For $n\geq 3$ the structure function $G_n$ fulfils
the weak time zero field condition \ref{1cond}.
\end{lemma}
\noindent{\bf Proof.}
By Parseval's theorem, the weak sharp time field condition for
$G_n$ is equivalent to the existence (in ${\cal S}_{n,j}'$) of the
limit
\begin{eqnarray}
\label{14eqa} &&\lim_{\epsilon\to+0}\int_{\R^{dn}} \hat
G_n(k_1,\ldots,k_n)\hat\varphi_{\epsilon}
(k_j^0)\hat\varphi_\epsilon(k_{j+1}^0)\nonumber\\
&\times&f(k_1,\ldots,k_{j-1},\vec k_j,\vec
k_{j+1},k_{j+2},\ldots,k_n)
 ~ dk_1\cdots dk_n
\end{eqnarray}
for $f\in{\cal S}_{n,j}$. We note that the norms $\|.\|_{1,d-1}$
on ${\cal S}_n$ in the proof of the temperedness of $\hat G_n$ in
Subsection 4.2 of \cite{AGW2} can be replaced by norms
$\|.\|_{1,d+1}'$ with $\|g\|_{K,L}', K,L\in\N_0,$ defined as
$$\sup_{ k_1,\ldots,k_n\in \R^{d}\atop 0\leq |\beta_1|,\ldots,|\beta_n|\leq K}\left|\prod_{l=j}^{j+1}(1+|k_l^0|^2)^{-{1\over 2}} [\prod_{l=1}^n(1+|\vec k_l|^2)^{L\over 2} {\partial^{|\beta_l|}\over (\partial k_l)^{\beta_l}}] g(k_1,\ldots,k_n)\right|
$$
$\forall g\in {\cal S}_n$ without changing the rest of the proof.
To see this, it is sufficient to check the simple estimate
$$
(1+|k_r^0|^2)^{1/2}\leq c \prod_{l=1}^n(1+|\vec
k_l|^2)^{1/2}~~\mbox{for } (k_1,\ldots,k_n)\in\mbox{supp } \hat
G_n
$$
where $r=1,\ldots,n$ and $c$ depends on $m_\kappa$,
$\kappa=1,\ldots,N$.

 Thus, $\hat G_n$ is continuous w.r.t. $\|.\|_{1,d+1}'$ (the argument in \cite{AGW2} is formulated only for a special case, but it carries over to the general case by a simple adaptation of notations, cf. \cite{Go}).

From the definition of $\varphi_\epsilon$ we get that $\hat
\varphi_\epsilon(x)=\hat\varphi(\epsilon x)$ and
$\hat\varphi(0)=1/(2\pi)^{1/2}$. From these properties we get that
the product of the two $\hat \varphi_\epsilon$ and $f$ in
(\ref{14eqa}) converges to $f(k_1,\ldots,k_{j-1},\vec k_j,\vec
k_{j+1},k_{j+2},\ldots,k_n)/(2\pi)$ w.r.t the topology induced by
$\|.\|_{1,d+1}'$ and thus the limit in (\ref{14eqa}) exists by the
continuity of $\hat G_n$ w.r.t. this norm. Furthermore, since
$f\to\|f\|_{1,d+1}'$ defines a Schwartz norm on ${\cal S}_{n,j}$,
the limit in (\ref{14eqa}) defines a tempered distribution in
${\cal S}'_{n,j}$.
\kasten

We now show the locality of the structure functions for the case
of Bosonic locality ($\sigma^{\kappa,\kappa'}=1$) by application
of Theorem \ref{1theo}.

\begin{theorem}
\label{2theo} The structure functions $G_n,~ n\geq 3$, are local.
\end{theorem}
\noindent {\bf Proof.}
By Theorem \ref{1theo} it suffices to show that $\hat
G_{n,[,]_j}(0,0)=0$, i.e.
\begin{equation}
\label{15eqa} \int_{\R}\int_{\R} \hat
G_n(k_1,\ldots,k_{j-1},[k_j,k_{j+1}],k_{j+2},\ldots,k_n)~dk_j^0dk_{j+1}^0=0~,
\end{equation}
where the double integral exists as a distribution in ${\cal
S}_{n,j}$ as a consequence of Lemma \ref{1lem}.

 For $j=1,\ldots,n-1$, the commutator in momentum space, $\hat G_{n,[,]_j}$, is
given by the following formula:
\begin{eqnarray}
\label{16eqa} &&\hat
G_n(k_1,\ldots,k_{j-1},[k_j,k_{j+1}],k_{j+2},\ldots,k_n)
=\left\{ \prod_{l=1}^{j-1}\delta_{\bar m}^-(k_l)\right.\nonumber\\
&\times&\sum_{\kappa_j,\kappa_{j+1}=1}^N
\bigg[{\delta_{m_{\kappa_j}}^-(k_j)\over
k_{j+1}^2-m_{\kappa_{j+1}}^2}+
{\delta_{m_{\kappa_{j+1}}}^+(k_{j+1})\over k_j^2-m_{\kappa_j}^2}-
{\delta_{m_{\kappa_{j}}}^-(k_{j+1})\over
k_{j}^2-m_{\kappa_{j+1}}^2}
\nonumber\\
&-&{\delta_{m_{\kappa_{j+1}}}^+(k_{j})\over
k_{j+1}^2-m_{\kappa_{j}}^2} \bigg]\left.
\prod_{l=j+2}^{n}\delta_{\bar
m}^+(k_l)\right\}\delta(\sum_{l=1}^nk_l)
\end{eqnarray}
Here we have used the notation $\delta^\pm_{\bar
m}=\sum_{\kappa_l=1}^N\delta_{m_{\kappa_l}}^\pm$. Changing the
order of summation in (\ref{16eqa}), we can replace the expression
in the brackets $[\ldots]$ by
$$
\left[{\delta_{m_{\kappa_j}}^-(k_j)\over
k_{j+1}^2-m_{\kappa_{j+1}}^2}+
{\delta_{m_{\kappa_{j+1}}}^+(k_{j+1})\over k_j^2-m_{\kappa_j}^2}-
{\delta_{m_{\kappa_{j+1}}}^-(k_{j+1})\over
k_{j}^2-m_{\kappa_{j}}^2}- {\delta_{m_{\kappa_{j}}}^+(k_{j})\over
k_{j+1}^2-m_{\kappa_{j+1}}^2} \right]
$$
We can now evaluate the integrals in (\ref{15eqa}) in the
following way: First the delta distributions
$\delta^{\pm}_{m_{\kappa_j}}(k_j)$
($\delta^\pm_{m_{\kappa_{j+1}}}(k_{j+1})$ ) are being used to
evaluate out the integration over $k_j^0$ (over $k_{j+1}^0$).
Then, we use the delta distribution
$\delta(\sum_{l=1}^{j-1}k_l^0\pm
\omega_{j,\kappa_j}+k_{j+1}^0+\sum_{l=j+2}^nk_l^0)$
($\delta(\sum_{l=1}^{j-1}k_l^0+k_j^0\pm\omega_{j+1,\kappa_{j+1}}+\sum_{l=j+2}^nk_l^0)$)
to evaluate the integral over $k_{j+1}^0$ ($k_j^0$, respectively)
where $\omega_{l,\kappa_l}=\sqrt{|\vec k_l|^2+m^2_{\kappa_l}}$. As
the result we get
$$
\left\{ \prod_{l=1}^{j-1}\delta_{\bar m}^-(k_l)
\sum_{\kappa_j,\kappa_{j+1}=1}^N \bigg[ ~~ \ldots ~~\bigg]
 \prod_{l=j+2}^{n}\delta_{\bar m}^+(k_l)\right\}\delta(\sum_{l=1}^n\vec k_l) ~
$$
with the expression $[\ldots]$ given by
\begin{eqnarray*}
&&\Bigg[ {1\over 2\omega_{j+1,\kappa_{j+1}}\left( (\omega_{j+1,\kappa_{j+1}}+a)^2-\omega_{j,\kappa_j}^2\right)}\\
&&+{1\over 2\omega_{j,\kappa_{j}}\left(
(\omega_{j,\kappa_{j}}-a)^2-\omega_{j+1,\kappa_{j+1}}^2\right)}-
\stackrel{j,j+1}{\longleftrightarrow}\Bigg],
\end{eqnarray*}
with $a=\sum_{l=1,l\not=j,j+1}^nk_l^0$. Here the notation
$\stackrel{j,j+1}{\longleftrightarrow}$ symbolizes, that the terms
standing before the arrow are being repeated with $j$ replaced by
$j+1$ and vice versa.

We let $x=\omega_{j,\kappa_{j}},~y=\omega_{j+1,\kappa_{j+1}}$ and
we want to show that
\begin{equation}
\label{20eqa} {1\over 2y\left( (y+a)^2-x^2\right)} +{1\over
2x\left( (x-a)^2-y^2\right)}-
\stackrel{x,y}{\longleftrightarrow}=0
\end{equation}
for $x,y,a\in \R$. We get for the left hand side of (\ref{20eqa}):
\begin{eqnarray*}
&&{1\over 2y(y+x+a)(y-x+a)}+{1\over 2x (x+y-a)(x-y-a)} -\stackrel{x,y}{\longleftrightarrow}\\
&=& {1\over 2xy(y-x+a)}\left( {x\over x+y+a}-{y\over x+y-a}\right) -\stackrel{x,y}{\longleftrightarrow}\\
&=&{1\over 2xy(y-x+a)}\left( {x(x-a)-y(y+a)\over
(x+y)^2-a^2}\right)-\stackrel{x,y}{\longleftrightarrow}.
\end{eqnarray*}
Since $1/(2xy((x+y)^2-a^2))$ is symmetric in $x$ and $y$, it
remains to show that
$$
{x(x-a)-y(y+a)\over y-x+a}-\stackrel{x,y}{\longleftrightarrow}=0.
$$
This is equivalent to
$$
(x-y+a)\left[
x^2-xa-y^2-ya\right]-\stackrel{x,y}{\longleftrightarrow}=0.
$$
Carrying out the multiplication we get
\begin{eqnarray*}
&&(\underbrace{x^3}_1-\underbrace{x^2a}_2-\underbrace{xy^2}_3-\underbrace{xya}_4)+(-\underbrace{yx^2}_3+\underbrace{xya}_4+\underbrace{y^3}_1+\underbrace{y^2a}_5)\\
&+&(\underbrace{x^2a}_2-\underbrace{xa^2}_6-\underbrace{y^2a}_5-\underbrace{ya^2}_6)-\stackrel{x,y}{\longleftrightarrow}=0,
\end{eqnarray*}
where we have labeled the terms which cancel each other or which
together give a expression symmetric in $x$ and $y$ with the
numbers $1$ to $6$. Thus, the above equation holds and the proof
is finished.
\kasten

\begin{remark}
\label{3.1rem} {\rm (i) In the definition of the structure
functions we can multiply the distributions
$\delta^\pm_{m_\kappa}$ and $1/(k^2-m^2_{\kappa})$ by a weight
$\lambda_\kappa\in\C$ and we obtain the locality of the
``weighted'' structure functions by the same arguments as used in
the proof of Theorem \ref{2theo}. By an approximation of the
integral by Riemannian sums and making use of the fact that the
limit of local distributions is again local, one immediately
obtains the locality of the distributions
\begin{eqnarray}
&&\int_0^\infty\!\!\cdots\int_0^\infty \!\!\bigg\{\sum_{j=1}^n\prod_{l=1}^{j-1}\delta^-_{m_l}(k_l) {1\over k^2_j-m_{j}^2}\nonumber \\
&\times&\prod_{l=j+1}^n\delta^+_{m_l}(k_l)\bigg\}\delta
(\sum_{l=1}^nk_l)~ \rho(dm^2_1)\cdots\rho(dm^2_n)
\end{eqnarray}
for a (sufficiently regular) locally finite and polynomially
bounded complex measure $\rho$ and, in a second step of
approximation, even for more irregular distributions $\rho$. This
re-establishes (and even slightly generalises) the result of
Theorem 7.10 of \cite{AGW1}.

\noindent (ii) In the proof of the locality of the two point
function of the free field and in Theorem \ref{2theo} the
polynomial in $\vec q_-$ was the simplest polynomial, i.e. the
zero function. This made it particularly simple to apply the
criterion Theorem \ref{1theo}. One can expect an analogous result
for all those Wightman functions in momentum space, where the
value of the Wightman function falls to zero whenever the
difference of two subsequently following momenta gets very large.
Such a behaviour can be justified in a number of physical
situations where the impact due to interaction declines if the
difference of momenta gets very large. }
\end{remark}

\section{Theorem 2.1 and the Jost-Lehmann-Dyson representation}
In this section we briefly compare the characterisation of
locality in momentum space given in Theorem 2.1 with the integral
representation of causal commutators by Jost, Lehmann and Dyson
(JLD): In \cite{Dy,JL} JLD consider matrix elements of the form
\begin{equation}
\label{4.1eqa}
f_{\Psi_1,\Psi_2}(\xi_-)=i\langle\Psi_1,[\phi(-\xi_-/2),\phi(\xi_-/2)]\Psi_2\rangle
\end{equation}
where $\Psi_1,\Psi_2$ are vectors\footnote{In \cite{Dy,JL} these
vectors are chosen to be (improper) eigen vectors of the
energy-momentum operator  $P$, however this is not important in
the present context.} from the standard domain ${\cal D}$ in the
representation Hilbert space ${\cal H}$ of the Wightman quantum
field (see \cite{SW}) $\phi(x)$. For simplicity we only consider
the case when $\phi(x)$ is a Bosonic, Hermitean and scalar field.
Also, we assume that the above expression exists as a distribution
in $\xi_-$ which is essentially equivalent with the assumption
that $\langle \Psi_1,[\phi(x),\phi(y)]\Psi_2\rangle$ is a function
in $\xi_+=(x+y)/2$ s.t. one can set $\xi_+=0$. Furthermore,
\cite{Dy,JL} require the decomposability of
$f_{\Psi_1,\Psi_2}(\xi_-)$ into an advanced and retarded part.
This is essentially the weak time zero field condition
\ref{1cond}. We do not want to enter into technical details and we
assume that $f_{\Psi_1,\Psi_2}(\xi_-)$ is a function in $\xi_-^0$
$\forall \Psi_1,\Psi_2\in{\cal D}$ (in the sense of Cond
\ref{1cond}) when smeared out in $\vec \xi_-$. Let $\hat
f_{\Psi_1,\Psi_2}(q_-)$ be the Fourier-transform of
$f_{\Psi_1,\Psi_2}(\xi_-)$. Then, Theorem \ref{1theo} takes the
following form:

\begin{corollary}
\label{1cor} Let the above assumptions be fulfiled. Then $\phi(x)$
is local if and only if $\int_{\R}\hat
f_{\Psi_1,\Psi_2}(q_-)dq^0_-$ is a polynomial in $\vec q_-$
$\forall \Psi_1,\Psi_2\in{\cal D}$.
\end{corollary}

In Eq. (12) of \cite{Dy} the JLD--representation is given for the
Fourier transform of the matrix element (\ref{4.1eqa}) of a local
quantum field $\phi(x)$:
\begin{eqnarray}
\label{4.2eqa}
\hat f_{\Psi_1,\Psi_2}(q_-)&=&\int_{\R^3}\int_0^{\infty} \varepsilon(q^0)\delta((q^0_-)^2-(\vec q_--\vec u)^2-\kappa^2)\nonumber\\
&\times& [\Phi_1(\vec u,\kappa^2)+q^0_-\Phi_2(\vec
u,\kappa^2)]~d\kappa^2d\vec u
\end{eqnarray}
where $\varepsilon(q^0_-)=\mbox{sign}(q^0_-)$ and $\Phi_1,\Phi_2$
are uniquely determined (generalised)  functions with support
properties depending on the spectrum of $\Psi_1,\Psi_2$. We
furthermore assume that $\Phi_1$ and $\Phi_2$ are sufficiently
integrable in order to make shure that the above representation
exists and that $f_{\Psi_1,\Psi_2}(\xi_-)$ fulfils the weak time
zero field condition. We then get by straight forward
calculations:
\begin{equation}
\label{4.3eqa} \int_{\R} \hat
f_{\Psi_1,\Psi_2}(q_-)~dq^0_-=\int_{\R^3}\int_0^\infty \Phi_2
(\vec u,\kappa^2)~d\kappa^2d\vec u=C,
\end{equation}
where $C$ is a constant and $C=0$ for an antisymmetric Bosonic
commutator. We thus see that locality of the JLD--representation
(\ref{4.2eqa}) is described by Corollary \ref{1cor} with the
additional restriction that all polynomials in $\vec q_-$ are
zero. This can be considered to be physically sufficient, cf.
Remark \ref{3.1rem} (ii), but it does not exhaust all mathematical
cases, as we shall explain using the structure functions $G_n$,
$n\geq 4$, of Section 3 for the simplest case where $N=1$ and
$m_\kappa=m>0$:

Let $M_n=M_n(k_1,\ldots,k_n)$ be a fully Lorentz invariant,
symmetric (under exchange of arguments)  and real polynomial. It
is easy to verify that $\hat W_n^T=M_n\cdot \hat G_n$ also is the
Fourier transform of a local Witghtman distribution, cf.
\cite{AG}. It is also easy to modify Lemma \ref{1lem} and to
verify the weak time zero field condition for all such $\hat
W_n^T$. It can be shown by explicit, but lengthy calculations that
the right hand side of Eq. (\ref{15eqa}) with $\hat G_n$ replaced
by $\hat W_n^T$ can not be equal to zero for some $M_n$ with
sufficiently high degree in any of the variables $k_l$ ($\geq 4$
is required at least).

Avoiding such tiresome calculations we give the following abstract
argument: Suppose that there is a JLD--representation (with
properties specified above) for each such $\hat W_n^T$. By Eq.
(\ref{4.3eqa}) the right hand side of Eq. (\ref{15eqa}) with $\hat
G_n$ replaced by $\hat W_{n,[,]_j}^T$ would be zero for all $M_n$.
Furthermore, fixing momenta $k_1,\ldots,k_{j-1},\vec k_j,\vec
k_{j+1},k_{j+1},\ldots,k_n$ to compact sets, we see that also
$k_j^0,k_{j+1}^0$ only run over compact sets. Thus we can
approximate in Eq. (\ref{15eqa}) a $C^\infty$ but non-analytic
function $\tilde M_n$ by polynomials $M_n$ on such compact sets.
Taking the limit, we see that also the r.h.s. of Eq. (\ref{15eqa})
would vanish for $\hat G_n$ replaced with $\hat
W_n^{T\infty}=\tilde M_n\cdot \hat G_n$. Then, by Theorem 2.1,
$\hat W_n^{T\infty}$ would be local and by [1, Theorem 4.5]
$\tilde M_n$ taken on-shell would be the (truncated) scattering
amplitude of a local relativistic quantum field theory (with
indefinite metric). This, however, is in contradiction with
crossing-symmetry, see e.g. \cite{Ep} where analyticity on certain
on-shell regions has been proved\footnote{I am grateful to D.
Buchholz for pointing out to me that crossing symmetry is being
violated by such approximation arguments.}. Thus, the
JLD--representation can not hold for all causal commutators of
$\hat W_n^T$ with arbitrary polynomial multiplier $M_n$.

The reason why the JLD--representation tacitly rules out some
cases of causal commutators is the following: In \cite{Dy} p. 1461
it is required that the distributional product
$f(\xi_-)\delta(\xi^2_--|y|^2)$ exists where $y\in\R^2$. Using
Eqs. (18)--(20) of \cite{Dy} it is easy to see that there are some
distributions $f(\xi_-)$ vanishing for $\xi_-^2<0$ s.t. the above
distributional product does not exist, take e.g. $d=4$,
$f(\xi_-)=\delta'(\xi_-^1)\delta'(\xi_-^2)\delta'(\xi_-^3)$ which
is constant in $\xi_-^0$ and thus fulfils the weak time zero field
condition. As we have demonstrated, at least from a mathematical
point of view, also such cases should be taken into account in
order to obtain a (more) complete characterisation of locality in
momentum space.

{\bf Acknowledgements} I would like to thank S. Albeverio for his scientific advice and a referee for pointing out to me references \cite{Dy,JL}.  This work was supported by DFG SFB 237 and DAAD ``Hochschulsonderprogramm III'' .


\begin{thebibliography}{99}
\bibitem{AG} S. Albeverio, H. Gottschalk, {\em Scattering theory for quantum fields with indefinite metric}, Univ. Roma 1 preprint 21/99, 1999.
\bibitem{AGW1} S. Albeverio, H. Gottschalk, J.-L. Wu,{\em Rev. Math Phys.}, Vol {\bf 8-6}, p. 763, (1996).
\bibitem{AGW2} S. Albeverio, H. Gottschalk, J.-L. Wu, {\em Commun. Math. Phys.} {\bf 184}, p. 509, (1997).
\bibitem{AGW3} S. Albeverio, H. Gottschalk, J.-L. Wu, {\em  Phys. Lett.} {\bf B 405}, p. 243 (1997).
\bibitem{Co} F. Contantinescu, {\em Distributionen und ihre Anwendung in der Physik}, Teubner, Stuttgart, 1973.
\bibitem{Dy} F. Dyson, {\em Phys. Rev.} {\bf 110-6}, p. 1460 (1958).
\bibitem{Ep} H. Epstein, {\em Some analytic properties of scattering amplitudes in quantum field theory}, in Axiomatic quantum field theory, Proc. 1965 Brandeis University Summer Scool on Theoret. Phys. Ed. M. Chretier, S. Deser, Gordon and Breach, New York 1966.
\bibitem{Jo} R. Jost, {\it General theory of quantized fields}, AMS Publications, 1963.
\bibitem{JL} R. Jost, H. Lehmann, {\em Nuovo Cimento} {\bf 5-6}, p. 1598 (1957).
\bibitem{Go} H. Gottschalk, Dissertation, Bochum 1999.
\bibitem{SW} R.F. Streater, A.S. Wightman, {\em PCT, spin and statistics, and all that}, Benjamin, New York, Amsterdam, 1964.
\end{thebibliography}
\end{document}